\documentclass[12pt]{article}
\usepackage{amssymb}
\usepackage{epsfig}

\usepackage{amsfonts,epsfig}

\def\pa{\partial}

\def\k{\kappa}

\def\d{\delta}
\def\r{\rho}

\def\l{\label}

\def\m{\mu}

\def\be{\begin{equation}}
\def\ee{\end{equation}}
\def\ba{\begin{eqnarray}}
\def\ea{\end{eqnarray}}

\def\t{\tilde}
\def\f{\frac}

\def\na{\nabla}

\begin{document}

\vspace{4cm}
\centerline{\large\bf Microscopic entropy of trapping horizon}
\vspace{1cm}
\centerline{\bf Mikhail Z. Iofa \footnote{ e-mail:
iofa@theory.sinp.msu.ru}} \centerline{Skobeltsyn Institute of Nuclear
Physics}
\centerline{Moscow State University}
\centerline{Moscow 119991, Russia}

\begin{abstract}
In the Carlip-Majhi-Padmanabhan approach,
we calculate microscopic entropy of trapping (apparent) horizon of the FRW metric.
We solve Killing equations for $t,r$ part of the metric without fixing {\it a piori}
the form of the scaling factor $a(t)$ which is determined from the
 requirement of consistency of Killing equations.
Further restrictions on the form of the Killing vector follow from the requirement
 that Killing vector is null  at the trapping horizon
at all $t$.
Applying the technique used to calculate
microscopic entropy of Killing horizons, we calculate  microscopic entropy of
trapping horizon. Using the explicit form of  Killing vector, we verify that identities
used in calculation of the central term of Virasoro algebra for Killing horizons of black holes
are valid in the present case. 

\end{abstract}

%%%%%%%%%%%%%%%%%%%%%%%%%%%%%%%%%%%%%%%%%%%%%%%%%%%%%%%%%%%%%%%%
\section{Introduction}
Since the discovery of thermodynamic nature of black holes
\cite{bek,haw}, calculation of
entropy of black holes has attracted considerable attention. On the one hand
thermodynamic properties of static event horizons of black holes were studied and their "geometric
entropy" was calculated.
Parallel to the study of black hole horizons and their thermodynamic properties
also cosmological horizons and their thermodynamics were intensively studied 
\cite{gibhaw,hayw4,haymuk,dev,cai,faraon}. In the course of these investigations it was realized that in  
cosmological cases more relevant is not the event horizon, but apparent 
and trapping horizons \cite{hayw4,asht,niels,nielsyeom,nielsviss}. 

On the other hand, much effort was
put to calculation of
"microscopic entropy". This approach initiated in \cite{micro}
originally in
2+1 gravity was subsequently developed in many papers. The approach is
based on the result \cite{Brown1} that in 2+1 gravity with negative
cosmological
constant the black hole possesses asymptotic symmetry
consisting of two copies of the Virasoro algebra. Substituting the
central charge
of the Virasoro algebra in the Cardy formula, it was possible to
calculate the density of states and to obtain  microscopic entropy
of black hole which appeared to be equal to the geometric one. Next, the approach based on the
use of the asymptotic symmetry, was generalized to 3+1 theory \cite{bar}
and applied to the Kerr black hole \cite{stro1}.

Another method to calculate microscopic entropy of black holes
was developed in papers of Carlip \cite{carlip1}, who showed
that symmetries
of near-horizon geometry imply the algebra of constraints with central
extension which contains the Virasoro sub-algebra with central
charge. The method was applied to 2D, 3D and 4D black holes \cite{carlip1,kang}. In
particular, in this approach were obtained the
results of
\cite{stro1} for Kerr/CFT \cite{carlip2}.
 Carlip's approach was refined in \cite{park}, and
\cite{carlip3},
with the use of new canonical hamiltonian which satisfies the variation
principle in presence of boundaries. Near-horizon geometry was investigated
in \cite{cvit}
The above results  were based on the on-shell constructions (see
review \cite{carlip3}).

Further development of the approach based on the use of the algebra of
diffeomorphisms in the near-horizon region was performed in papers
\cite{padm1}, where an off-shell construction of the algebra of
constraints was developed. The off-shell Noether current was used to calculate the
central charge of the algebra of diffeomorphisms and to obtain the Wald
entropy \cite{wald}. The method provides a  straightforward
derivation
of the Virasoro algebra and calculation of the central term.

The above results which refer to local Killing horizons 
 of stationary black holes, 
defined as a surface at which Killing vector is null,  
 were generalized to a
more general case of isolated horizons \cite{isol},
which are generalization of Killing horizon of black hole, and
which include non-stationary geometries. Imposing boundary conditions on the metric in the
near-horizon region, a
set of diffeomorphisms was constructed with the corresponding Noether
charges forming the Virasoro algebra.

In the study of metrics with dynamically evolving non-stationary horizons
 rather than stationary black holes, 
definition of the boundary of a stationary black hole
as a hypersurface at which Killing vector is null was changed to
definition of the boundary as a trapping horizon which is
 the surface where Kodama vector
 is null \cite{kodama,hayward1,hayward2}.
Thermodynamic of evolving horizons was studied by substitution of 
Kodama vector and trapping horizon for Killing vector and Killing horizon.

In this paper we study possibility to calculate microscopic 
entropy of trapping horizon of spherically-symmetric metrics 
$$
ds^2 = -dt^2 +a^2 (t)b^2 (r)dr^2+ R^2(r,t) d\Omega^2
$$
by reducing the problem to calculation through the null Killing vector.
For this class of metrics condition that Kodama vector
$K^i =(1/\sqrt{-g})\epsilon^{ij}\pa_j R$
is null can be alternatively written as $g^{ik}\pa_i R\pa_k R=0$ \cite{kodama,hayward1,hayward3}.
To perform a calculation of microscopic entropy of the trapping horizon,
 we look for a Killing vector of the metric $-dt^2 +a^2 (t)b^2 (r){dr^2}$.
Solving Killing equations we do not fix the scaling factor $a(t)$ 
which form is restricted by
 consistency conditions of the system of Killing equations.
 Further restrictions on the components of the Killing vector are obtained,
 by demanding that Killing vector is null  at the trapping horizon
for all $t$. 

Because by construction  Killing vector is null at the trapping horizon, 
Killing and trapping horizons coincide, and we
can apply the well-developed methods of calculation of microscopic entropy of
 local Killing horizons \cite{carlip1,carlip2,kang,cvit,padm1}.
 We use the off-shell approach of \cite{padm1}. 
We find that Killing vector null at the trapping (apparent) horizon
exists in the case  of the FRW metric with $b =1$ and $R^2=a^2 (t)r^2$. 
We prove that relations derived in \cite{carlip1}
and \cite{padm1} used in calculations of the central term of the Virasoro algebra are valid in the present 
case. We calculate  microscopic entropy
of the trapping horizon which
 coincides with the result of
thermodynamic calculations (reviews \cite{hayward1,padm2,faraon}).

We also consider the FRW metric with $b^2 (r)=(1\pm r^2 )^{-1}$ and find that 
in these cases Killing equations do not admit null Killing vectors on trapping horizon.

%%%%%%%%%%%%%%%%%%%%%%%%%%%%%%%%%%%%%%%%%%%%%%%%%%%%%%%%%%%%%%%%%%
\section{Killing vectors on trapping horizon}

We consider a spatially homogeneous isotropic space-time with the metric
\be
\l{2.1}
ds^2 =g_{ij}dx^i dx^j +R^2 (x) d\Omega^2 =-dt^2 +a^2 (t)
\f{dr^2}{1-kr^2}+a^ 2 (t)r^2 d\Omega^2,
\ee
where $k=\pm 1,0$. We begin with the metric with $k=0$.
 Killing equations $L_\chi g_{ij}=0 $ have the form
\ba
\l{2.2}
\l{2a}
&{}& \chi^c \pa_c g_{tt} + 2\pa_t \chi^c g_{ct}=0,\\
\l{2b}
&{}& \chi^c \pa_c g_{rr} + 2\pa_r \chi^c g_{cr}=0,\\
\l{2c}
&{}& \chi^c \pa_c g_{rt} + \pa_t \chi^c g_{cr}+\pa_r \chi^c g_{ct}=0.
\ea
Substituting explicit expressions for the metric, we reduce Killing equations to a system
\ba
\l{3a}
&{}&\dot {\chi}^t =0,\\         
\l{3b}
&{}&2(\chi^r)' a^2+ \chi^t (\dot{a^2}) =0,\\         
\l{3c}
&{}&-(\chi^t)' +\dot{\chi}^r {a}^2 =0,
\ea
where  dot and prime are derivatives over $t$ and $r$.
Integrating the system, we obtain
\ba
&{}&\l{2.4} \chi^t = f(r), \\
&{}&\l{2.5} \chi^r=-\f{\dot{a}}{a}\int^r f(r')dr'+\varphi (t),\\
&{}&\l{2.6} \chi^r = f'(r) \int^t  \f{dt'}{a^2 (t')}+\psi (r).
\ea
where $f,\,\,\,\varphi$ and $\psi$ are arbitrary functions.
Taking derivatives of (\ref{2.5}) and (\ref{2.6}) over $r$ and $t$
and equating the resulting expressions, we obtain
\be
\l{2.7}
f''(r) + (\ddot{a}(t) a(t) -\dot{a}^2 (t)) f(r) =0.
\ee
For consistency of Eq. (\ref{2.7}) it is necessary that $\ddot{a} a(t)
-\dot{a}^2 (t) =const$.

%%%%%%%%%%%%%%%%%%%%%%%%%%%%%%%%%%%%%%%%%%%%%%%%%%%%%%%%%%%%%%%%%%%%%%%%%
\subsection{The case $\ddot{a} a-\dot{a}^2 =0$}
%%%%%%%%%%%%%%%%%%%%%%%%%%%%%%%%%%%%%%%%%%%%%%%%%%%%%%%%%%%%%%%%%%%

First we consider the case $const =0$. Equation $\ddot{a} a -\dot{a}^2 =0$
has  solution
\be
\l{2.8}
a(t)= C_1 e^{-Ht}
,\ee
where $H$ and $C_1$ are integration constants.
From (\ref{2.7}) we have
\be
\l{2.81}
  f(r) = k_1 +k_2 r
\ee
Eqs. (\ref{2.4}) - (\ref{2.6}) yield
\ba
\l{2.9} &{}&\chi^t =k_1 +k_2 r\\
\l{2.10a}&{}&\chi^r = H\left(k_1 r +k_2\f{r^2}{2} +C' \right) +\varphi (t)\\
\l{2.10b}
&{}&\chi^r= k_2 \left(\f{e^{ 2Ht}}{ 2HC_1^2} +C''\right) +\psi (r).
\ea
Equating expressions (\ref{2.10a}) and (\ref{2.10b}) and noting that dependence on $r$ and $t$ is 
separated,
we have 
$$
\varphi (t)=\f{k_2}{2H\,a^2 (t)}+C_2,\qquad\,\, \psi (r)= H\left(k_1 r +k_2\f{r^2}{2}\right) +C_3
,$$ 
where $HC'+C_2 =k_2 C'' +C_3 \equiv C$. Substituting these relations in (\ref{2.10a}) and (\ref{2.10b}), we 
obtain
\be
\l{2.10}\chi^r = H\left(k_1 r +k_2\f{r^2}{2} \right)+ \f{k_2}{2H\,a^2}  +C
\ee
%%%%%%%%%%%%%%%%%%
Collecting (\ref{2.9}) and (\ref{2.10}), we have
\be
\l{2.11} 
\chi^a =\left( k_1 +k_2 r;\,\,\, H\left(k_1 r +k_2\f{r^2}{2} \right)+ \f{k_2}{2H\,a^2}  +C\right)
\ee
and
\be
\l{2.121}
\chi^2 =-{(k_1 +k_2 r)}^2 +a^2 H^2 \left(k_1 r +k_2\f{r^2}{2} + \f{k_2}{2H^2\,a^2}  +C\right)^2.
\ee
Dynamical trapping horizon is found by solving the equation
\be
\l{2.13}
g^{ik}\pa_i R\pa_k R =0.
\ee
Eq. (\ref{2.13}) yields
\be
\l{2.14}r_h =\f{1}{|\dot{a}(t)|}=\f{1}{Ha(t)}.
\ee
It  is seen that at the trapping horizon the Killing vector (\ref{2.11}) with $C=0$
\be
\l{2.15} \chi^a \big|_{hor}
=\left(k_1 +\f{k_2}{Ha};\,\,\,\f{1}{a}\left( k_1 + \f{k_2}{Ha}\right)\right)
\ee
is null.

There exist also a solution with $f(r)=0,\,\,\,\,\chi^a =(\chi^t , \chi^r )= (0, 1)$.
However,
 the norm of this Killing vector is $\chi^2 = a^2 (t)$ and does not vanish
on horizon.

%%%%%%%%%%%%%%%%%%%%%%%%%%%%%%%%%%%%%%%%%%%%%%%%%%%%%%%%%%%%%%%%%%%%%%%%%%%%%%%
\subsection{The case $\ddot{a} a-{\dot{a}}^2=\pm m^2$ }

Let us consider the case $const =-m^2$.
Equation $\ddot{a} a-{\dot{a}}^2=-m^2$ has solution
\be
\l{B.1}
a(t) =\f{m}{H}\sinh(Ht +p)
\ee
where $H$ and $p$ are arbitrary. Solution of the Eq. (\ref{2.7}), $f'' -m^2 f=0$, is
\be
\l{B.2}
f(r)=C_1 e^{mr} +C_2 e^{-mr}.
\ee
From Eqs.(\ref{2b}) and (\ref{2c}) we obtain
\be
\l{B.3}
\chi^r =-\f{H}{m^2}f' (r)(\coth(Ht+p) +C' )+\psi (r)
\ee
and
\be
\l{B.4}
\chi^r =-\f{H}{m^2}\coth(Ht+p)(f' (r) + C'' ) + \varphi (t),
\ee
where $\psi$ and $\varphi$ are arbitrary functions.
Equating (\ref{B.3}) and (\ref{B.4}), we obtain
\be
\l{B.5}
-\f{H}{m^2}f' (r)C' +\psi (r) = -\f{H}{m^2}\coth(Ht+p)C'' +\varphi (t) =C,
\ee
where
\be
\l{B.51}
\psi (r)=C+ \f{H}{m^2}f' (r)C';\qquad \varphi (t) =C +\f{H}{m^2}\coth(Ht+p)C''
\ee.
Substituting (\ref{B.51}), we find the Killing vector 
\be
\l{B.7}
\chi^a=\left(f(r);\,\,\, -\f{H}{m^2}f'(r)\coth(Ht+p) +C \right).
\ee
Horizon is located at 
$$r_h^{-1} ={m} \cosh (Ht +p).
$$ 
It is seen that condition $\chi^2_h =0$ is not satisfied for all $t$
\footnote{For example, taking in (\ref{B.2}) $C_2 =0$ from condition $\chi^2_h =0$ we obtain $\pm 
1=\f{H}{m}\coth(Ht+p)+C$.}.

The case  $\ddot{a}a -{\dot{a}}^2 =m^2$ is 
obtained by substitution $m\rightarrow im$ and leads to the same conclusion as above.

%%%%%%%%%%%%%%%%%%%%%%%%%%%%%%%%%%%%%%%%%%%%%%%%%%%%%%%%%%%%%%%%%%%%%%%%%%%
\subsection{Killing vectors for the metric with $k=\pm 1$}
%%%%%%%%%%%%%%%%%%%%%%%%%%%%%%%%%%%%%%%%%%%%%%%%%%%%%%%%%%%%%%%%%%%%%%%%%%%%%%

To be precise, we consider Killing equations for the metric with $k=-1$
\be
\l{A.1}
ds^2 =-dt^2 + dr^2\f{a^2 (t)}{1+r^2 }+a^2 (t) r^2 d\Omega^2
.\ee
Killing equations are
\ba
\l{A.2}
&{}&\dot {\chi}^t =0,\quad {\chi}^t\Rightarrow f(r)\\
\l{A.3}
&{}&\chi^r a^2 \f{-2r}{(1+r^2 )^2}+ \chi^t \f{(\dot{a^2})}{1+r^2}
+2(\chi^r)'\f{a^2}{1+r^2} =0,\\
\l{A.4}
&{}&-(\chi^t)' +\dot{\chi}^r \f{a^2}{1+r^2} =0,
\ea
or
\ba
\l{A.21}
&{}&(\chi^r)'+\f{\dot{a}}{a}f -\chi^r\f{r}{1+r^2}=0\\
\l{A.31}
&{}&\dot{\chi}^r =f'\f{1+r^2}{a^2}
\ea
Taking derivative of Eqs.(\ref{A.21}) and (\ref{A.31}) over $t$  and 
$r$ and equating the resulting expressions for $\dot{\chi}^{r'}$, we obtain  consistency equation
\be
\l{A.5}
 [f' (r)(1+r^2 )]'-f' (r) r+[\ddot{a}(t)a(t)-{\dot{a}(t)}^2 ] f(r)=0.
\ee
In the case $\ddot{a}a-{\dot{a}}^2 =0$ we obtain $a(t) =C_1 \exp{\{-Ht\}}$ 
 and 
\be
\l{a.51}
f(r)=k_1 +k_2 \ln[r+\sqrt{1+r^2}] 
,\ee
where $k_{1,2}$ are integration constants. 
Substituting $f(r)$ and $a(t)$ in Eqs. (\ref{A.3}) and (\ref{A.4}) and performing integration,
 we obtain
solutions of these equations
\ba
\l{A.6}
&{}&\chi^r =\sqrt{1+r^2}\left(\varphi (t) +H\int \,dr\f{f (r)}{\sqrt{1+r^2}}\right)
=\sqrt{1+r^2}\left(\varphi (t) +\f{f^2(r)H}{2k_2}+C'\right)\\
\l{A.7}
&{}&\chi^r = k_2\,\sqrt{1+r^2} \int \,\f{dt}{a^2 (t)}+\psi (r)=\sqrt{1+r^2}\left(\f{k_2}{2H\,a^2 (t)} 
+C''\right) +\psi (r) ,
\ea
where $\varphi (t)$ and $\psi (r)$ are arbitrary functions.
Equating expressions  for $\chi^r/\sqrt{1+r^2}$ which follow from (\ref{A.6}) and (\ref{A.7})
 and noting that  dependence on $r$ and $t$ is separated, we obtain
$$
C'+\varphi (t)  -\f{k_2}{2H\,a^2 (t)} = C'' + \f{\psi (r)}{\sqrt{1+r^2}} 
-\f{f^2(r)H}{2k_2} =C  
.$$
Killing vector is
\be
\l{A.8}
\chi^a= \left( f \,;\,\sqrt{1+r^2}\left(\f{H}{2k_2}f^2 +\f{k_2}{2H\,a^2} +C\right)\right).
\ee
Trapping horizon is located at 
\be
\l{A.1a}
r^2_h=(\dot{a}^2 -1)^{-1}.
\ee
Square of the Killing vector is
\be
\l{A.9}
\chi^2 =-f^2 +a^2\left(\f{H}{2k_2}f^2 +\f{k_2}{2H\,a^2} +C\right)^2.
\ee
Supposing that $\chi^2$  vanishes on horizon,  we  have
\be
\l{A.10}
f(r_h ) =\pm \f{k_2}{Ha} +\t{C} ,
\ee
or, using (\ref{a.51}),
\be
\l{A.11}
k_1 +\f{k_2}{2}\ln\f{Ha+1}{Ha-1} = \pm \f{k_2}{Ha} +\t{C}.
\ee
Relation (\ref{A.11}) is not valid for all $t$, and we conclude that $\chi^2$ 
does not vanish on the horizon for all $t$.

In the case $\ddot{a}a-{\dot{a}}^2 =-m^2$ we  solution for $a(t)$ and $f(r)$ is
\ba
\l{A.12}
a(t) =\f{m}{H}\sinh(Ht +p), \qquad
f(r)=k_1 p^m +k_2 p^{-m}
,\ea
where
$$
p= r+\sqrt{1+r^2}
.$$
Trapping horizon is at
\ba
\l{A.13}
r^2_h = (m^2 \coth^2 (Ht+p)-1)^{-1}
.\ea
 Killing vector has the following structure
\ba
\l{A.14}
\chi^a =\left(k_1 p^m +k_2 p^{-m};\,\,\, -\f{H}{m}\sqrt{1+r^2}(k_1 p^m -k_2 p^{-m})\coth (Ht+p)\right)
.\ea
 Assuming that the square of the Killing vector is null at the trapping horizon, we obtain a relation
$$
\sqrt{m^2 \coth^2 (Ht+p)-1}=m \coth^2 (Ht+p)
$$
which is not valid for all $t$.

To conclude, the metrics (\ref{2.1}) with $k\neq 0$ do not admit Killing vectors null
at the trapping horizon.

%%%%%%%%%%%%%%%%%%%%%%%%%%%%%%%%%%%%%%%%%%%%%%%%%%%%%%%%%%%%%%%%%%%%%%%%%%%
\section{Local Killing horizon}
%%%%%%%%%%%%%%%%%%%%%%%%%%%%%%%%%%%%%%%%%%%%%%%%%%%%%%%%%%%%%%%%%%%%%%%%%%%

In this section we introduce objects used  to construct the Virasoro 
algebra.
The surface gravity is defined as
\be
\l{3.1}
\k^2 =-\f{1}{2}\na^a \chi^b\na_a\chi_b|_{hor}.
\ee
For the Killing vector (\ref{2.10}) we obtain
\footnote{Because Killing vector $\chi^a$ is defined up to a constant factor, we can set
 $k_1 =1$.  In this
case the surface gravity (\ref{sg}) coincides with that defined in \cite{hayward1}.}
\be
\l{sg}
\k^2 =H^2 k^2_1.
\ee
Next, we define the vector $\r_a$
\be
\l{3.2}
\r_a =-\f{1}{2\k}\na_a \chi^2,
\ee
which explicitly is
\be
\l{3.3}
\r_a = \f{1}{k_1}\left( (aHF)^2 -\left(\f{k_2}{2Ha}\right)^2;
\,\,a\,F'\left(\f{k_2}{2Ha}- F\,Ha\right)   \right).
\ee
It is verified explicitly that $\chi^a\r_a =0$
\footnote{It is also valid on general grounds since $\chi^a$ is a Killing vector.}.
On the horizon we have
\be
\l{3.31}
\r_{a,hor}=\left(k_1 +\f{k_2}{Ha};\,\,\,-a\left(k_1
+\f{k_2}{Ha}\right)\right) .
\ee

Sufficient condition that surface variations keep position of horizon fixed is
\be
\l{3.4}
\f{\chi^a \chi^b \d g_{ab}}{\chi^2}\rightarrow 0 ,
\ee
 as $\chi^2 \rightarrow 0$.
Carlip considered diffeomorphisms of the form
\be
\l{3.5}
\xi^a =T\chi^a + R\r^a.
\ee
For these diffeomorphisms one has
\be
\l{3.6}
\d\chi^2 =\chi^a \chi^b \d_\xi g_{ab}= 2\chi^2 D T -2\k \r^2 R,
\ee
where $D=\chi^a \na_a$.
To form a closed algebra of diffeomorphisms near horizon
the functions $T$ and $R$ are constrained by additional conditions \cite{carlip1,padm1}
\footnote{It seems that in the off-shell approach condition $\int\, d^2 x\sqrt{h} D^3 T =0$ is unnecessary
due to cancellations.}.
\be
\l{3.7}
\r^a\na_a T=0, \qquad R=-\f{\chi^2}{\k\r^2}DT,  \qquad \na_a T=\f{\chi_a}{\chi^2}DT
\ee
In the near-horizon region the first condition (\ref{3.7}) reduces to
\be
\l{3.8}
\left(k_1 +\f{k_2}{Ha}\right)\left(\pa_t T(r,t) + \f{1}{a(t)}\pa_r T(r,t)\right)=0
\ee
with solution $T=T(r -1/aH)$.
 The second condition (\ref{3.7}) in the hear-horizon region is
$$
R=-\f{\chi^2}{\k\r^2}\left(k_1 +\f{k_2}{Ha}\right)\left(\pa_t T +\f{1}{a}\pa_r T\right)
.$$
 Because of (\ref{3.8}), $R$ is zero at the horizon.
In Appendix it is shown that the ratio $\chi^2/\r^2$ is finite at the horizon.

%%%%%%%%%%%%%%%%%%%%%%%%%%%%%%%%%%%%%%%%%
\section{Algebra Virasoro and Cardy formula}
%%%%%%%%%%%%%%%%%%%%%%%%%%%%%%%%%%%%%%%%%%%%%%%%%%%%%

In this section we briefly consider some steps of the Carlip-Majhi-Padmanabhan
approach
leading to the Virasoro algebra.
The formulas which were derived in \cite{carlip1,padm1} 
using only properties of Killing vectors require no further
justification,
validity of those
which were obtained relying on some properties of static horizons 
are discussed in Appendix. Below we follow notations of \cite{carlip1,padm1}.

The current associated with the diffeomorphism $\xi^a$ (\ref{3.5}) is given by
\be
\l{4.1} J^a =\f{1}{8\pi G}(\na^d\na^a\xi_d -\na^2 \xi^a )
\ee
In the vicinity of horizon the current takes the form
\be
\l{4.2}J^a =-\f{1}{8\pi G}\f{\r^a}{\chi^2}\left(2\k D T -\f{1}{\k}D^3 T\right) +O(\chi^2)
\ee
The Noether charge is corresponding to diffeomorphisn $\xi$ is
\ba
\l{4.3}
Q(\xi )=&{}&\int d\Sigma_a\sqrt{g}J^a =\int d\Sigma_{ab}\sqrt{h}J^a\xi^b
\\\nonumber
&{}&=-\f{1}{32\pi G}\int d^2 x \sqrt{h }\left|\f{\r}{\chi}\right|(g^{ac}g^{bd}-
g^{ad}g^{bc})\mu_{ab}\mu_{cd}\left(2\k DT -\f{1}{\k}D^3T\right)
\ea
Where $d\Sigma_{ab}=d^2 x\,\mu_{ab}$ is the surface element on the Killing horizon.
Constructing
\be
\l{4.4}d\Sigma_{ab}J^b (\xi_1 )\xi_2^a =\f{d^2 x}{32\pi G}\left|\f{\r}{\chi}\right|(g^{ac}g^{bd}-
g^{ad}g^{bc})\mu_{ab}\mu_{cd}\left(2\k DT_1 -\f{1}{\k}D^3T_1\right)T_2
\ee
one obtains commutator $[Q_1 , Q_2]=Q(\{\xi_1 ,\xi_2\})+K(\xi_1,\xi_2) $ where the central term
$K(\xi_1,\xi_2) $ in the near-horizon limit is
\be
\l{4.5}
K(\xi_1,\xi_2) =-\f{1}{32\pi G}\int d^2 x \sqrt{h }(g^{ac}g^{bd}-
g^{ad}g^{bc})\mu_{ab}\mu_{cd}\f{1}{\k}(DT_1 D^2 T_2 - DT_2 D^2 T_1 ).
\ee
Next, one defines decomposition of $T_{1,2}$ in the basis
\be
\l{4.6}
\{ T_n = \f{1}{\k}e^{in\k v +f(\tau^i )} \}
\ee
Here $f(\tau^i )$ is a function  of other coordinates regular on horizon
and $v$ is defined as $\chi^a \na_a v=1$
\footnote{In the near-horizon region  $v=v(r -1/a(t)H)$.} .
 It follows that
\be
\l{4.7}
\{T_m , T_n \}=T_m DT_n - T_n DT_m =-i(m-n )T_{m+n}
.\ee
Using (\ref{4.3})-(\ref{4.4}) it is shown that
\be
\l{4.8}
i[Q(\xi_m ), Q(\xi_n )]=(m-n)Q(\xi_{m+n}) +m^3 \f{A}{8\pi G}\d_{m+n, \,0}
\ee
where $A$ is the horizon area.

Defining the central charge and zero mode eigenvalue, and substituting them in the Cardy
formula, one obtains the entropy equal to the geometric one, i.e.$ S=A/4G$.

%%%%%%%%%%%%%%%%%%%%%%%%%%%%%%%%%%%%%%%%%%%%%%%%%%%%%%%%%%%%%%%%%%%%%%%%%%%%%%%

\section{Conclusions}

%%%%%%%%%%%%%%%%%%%%%%%%%%%%%%%%%%%%%
In this paper we calculated microscopic entropy of trapping horizon of the FRW metric
by looking for a null Killing vector at
the trapping horizon of the 2D $r,t$ part of the metric. 
Having a null Killing vector, we can apply the well-developed technique of
calculation of microscopic entropy of null surfaces.
Killing equations are solved
without {\it a priori} fixing the form of the scaling factor $a(t)$, which is
determined from consistency condition of Killing equations  $\ddot{a}a-{\dot{a}}^2 =const$. 
We have found that
in the case $const =0$ the FRW metric (\ref{2.1}) with $k=0$ has Killing vector null
at the trapping horizon.
In this  case the metric is dS metric.
The metrics (\ref{2.1}) with $k=\pm 1$  do not
admit Killing vectors null at the trapping horizon. 

The metric (\ref{2.1}) with $k=0$ is De Sitter metric in the "flat slicing". In
static coordinates De Sitter metric is
$$
ds^2 = -\left(1-\f{r^2}{H^2}\right)dt^2 + \left(1-\f{r^2}{H^2}\right)^{-1} dr^2
+r^2 d\Omega^2
.$$
In this form the metric has static event horizon 
 and  belongs to a  class of
metrics discussed in
\cite{carlip1,carlip2,padm1}, and one can apply the methods
of these papers to obtain Virasoro
algebra of surface deformations \cite{lin}.

It is of interest to find further examples of metrics with trapping (apparent) horizons to which one
can apply the above
procedure. However it is not an easy job to find a null Killing vector for a
general metric.
A minor generalization of the above metrics is
$$
ds^2 = -dt^2 +a^2 (t)b^2 (r)dr^2 +R^2(x) d\Omega^2
,$$
where $a(t)$ and $b(r)$ are not fixed {\it a priori}. 
Transforming $b(r)dr= d\r$ we obtain the metric with $b=1$ and transformed $\t{R}(x)$.
Killing equations can be solved as in Sect.2, and $a(t)$ is determined from 
the equation  $\ddot{a}a-{\dot{a}}^2 =0$. 

Demanding that at the horizon the Killing vector is null, we arrive at the equation
$$
{(k_1 +k_2 \r_h)}^2 =a^2 H^2 \left(k_1 \r_h +k_2\f{\r_h^2}{2} + \f{k_2}{2H^2\,a^2}  +C\right)^2
,$$
or, taking  the square root of both parts of the relation with the sign plus,
$$
k_1\left(\f{1}{Ha}-\r_h \right)=\f{k_2}{2}\left(\f{1}{Ha}-\r_h \right)^2 +C
.$$
From this equation we obtain
$$
\r_h=\f{1}{Ha}-\f{k_1}{k_2}\pm\sqrt{\f{k_1^2}{k_2^2}-C}= \f{1}{Ha}+\t{C}.
$$
To obtain a tractable form of equation for trapping horizon,
 -$\dot{\t{R}}^2 +a^{-2}{\t{R'}}^2 =0$,  we are forced again to assume
a separable form of $\t{R}(x)$. 
The above form of $\r_h$ is obtained by taking $\t{R}=a(t)(\r-\t{C})C'$. Transforming back
to variable $r$, we obtain $R(r,t)=a(t)(\int\,b(r)dr -\t{C})C'$.

%%%%%%%%%%%%%%%%%%%%%%%%%%%%%%%%%%%%%%%%%%%%%%%%%%%%%%%%%%%%%%%%%%%%%%%%%%%%%%%%%%%%%%%%   
\section{Appendix }
\renewcommand{\theequation}{A.\arabic{equation}}
\setcounter{equation}{0}  % reset counter

In this Appendix we explicitly verify some relations
used in derivation of the central extension of the Virasoro algebra.
We consider the case discussed in Sect. 2.1.

Let us denote by $x$ variation of the radius of the trapping horizon $r_h =1/\dot{a}(t):\,\,\,
r=r_h +x$.  Under this variation the vectors $\chi^a$ and $\r_a$ change as
\be
\l{C.1}
\chi^a_h+\d \chi^a =\left(  \left(k_1 +\f{k_2}{Ha}\right)
+k_2 x;\,\,\,\f{1}{a}\left(k_1+\f{k_2}{Ha}\right)+
xH\left(k_1+\f{k_2}{Ha}\right) \right)
\ee
\ba
\l{C.2}
\r_{a,h} +\d\r_a= \left(   \left(k_1+\f{k_2}{Ha}\right)
+x\f{2Ha}{k_1}\left(k_1+\f{k_2}{Ha}\right) \left(k_1+\f{k_2}{2Ha}\right);\,\,\,\right.
\\
\left. -a\left (k_1+\f{k_2}{Ha}\right)-x\f{ Ha^2}{k_1}\left(k^2_1 +\f{3k_1 k_2}{H a}
+\f{k_2^2}{H^2 a^2}\right)  \right)
\ea
It is explicitly verified that in the first order in $x\,\,\,
\chi^a \r_a =0$.
 From the expressions (\ref{C.1}), (\ref{C.2}) in the first order in $x$ we obtain
\ba
\l{C.6}
&{}&(\chi_h +\d \chi )^2 =2xk_1 {Ha}\left(k_1+\f{k_2}{Ha}\right),\\
\l{C.6a}
&{}&(\r_h +\d\r )^2=-2xk_1 Ha \left(k_1+\f{k_2}{Ha}\right).
\ea
Here $\chi^2_h$ and $\r^2_h$=0.
From the formulas  (\ref{C.1})-(\ref{C.3}) it
 follows that $|\chi/\r|\rightarrow 1 $ near horizon.
It also follows that the vector
$$k^a=\f{1}{\chi^2}\left(\chi^a -|\f{\xi}{\r}|\r^a\right)
$$
is finite on the horizon.

From (\ref{C.1}), (\ref{C.2}) we obtain
\be
\l{C.3}
\chi_r\chi_t -\chi_t\chi_r =2x\,k_1 Ha^2 \left(k_1 +\f{k_2}{Ha}\right)
\ee
Calculating
\be
\l{C.4}
\na_r\chi_t =-\na_t\chi_r =Ha\,k_1
,\ee
from the formulas (\ref{C.3}), (\ref{C.4}) and (\ref{C.6}) we have an important relation
\be
\l{C.5}
\na_a\chi_b = \f{\k}{\chi^2}(\chi_a\r_b -\chi_b\r_a )
.\ee
In the same way, calculating $\chi_a\chi_b -\r_a\r_b$ (as an example we present result for
$a,\,b =r,\,t)$
\be
\l{C.7}
\chi_r\chi_t -\r_r\r_t =2xa^2H \left(k_1 +\f{k_2}{Ha}\right)^3
\ee
and
\be
\l{C.8}
\na_r\chi_t =\na_t\chi_r =\f{Ha}{k_1} \left(k_1 +\f{k_2}{Ha}\right)^2
,\ee
we obtain
\be
\l{C.9}
\na_a\r_b =\f{\k}{\chi^2}((\chi_a\r_b -\chi_b\r_a ).
\ee
Relations (\ref{C.5}) and (\ref{C.9}) are the main building blocks leading to the formula  (\ref{4.2})
It follows also that the surface element
\be
\l{C.10}
\m_{ab} =-\bigg|\f{\chi}{\r}\bigg|(\chi_a\r_b -\chi_b\r_a )
\ee
is finite at the horizon and  $\m_{rt}=a(t)$.
%%%%%%%%%%%%%%%%%%%%%%%%%%%%%%%%%%%%%%%%%%%%%%%%%%%%%%%%%%%%%%%%%%%%%%%%%%%%%%%%

\end{document}